\documentclass[
twocolumn
,floatfix
]{revtex4}

\usepackage{amsmath}
\usepackage{amssymb}
\usepackage{graphicx}

\begin{document}

\title{A certification scheme for the boson sampler}

\newcommand{\affila}{National Key Laboratory of Electromagnetic Environment, China Research Institute of Radiowave Propagation, Qingdao, 266107, China}
\newcommand{\affilb}{Department of Physics, Ocean University of China, Qingdao 266100, China}
\newcommand{\affilc}{Centre for Quantum Computation and Communication Technology, School of Mathematics and Physics, University of Queensland, St Lucia, Queensland 4072, Australia}

\author{Kai Liu}
\affiliation{\affila}
\affiliation{\affilb}
\author{Austin Peter Lund}
\affiliation{\affilc}
\author{Yong-Jian Gu}
\affiliation{\affilb}
\email[Corresponding author:]{guyj@ouc.edu.cn}
\author{Timothy Cameron Ralph}
\affiliation{\affilc}



\begin{abstract}
Boson sampling can provide strong evidence that the computational power of a quantum computer outperforms a classical one via currently feasible linear optics experiments. However, how to identify an actual boson sampling device against any classical computing imposters is an ambiguous problem due to the computational complexity class in which boson sampling lies. The certification protocol based on bosonic bunching fails to rule out the so-called mean-field sampling. We propose a certification scheme to distinguish the boson sampling from the mean-field sampling for any random scattering matrices chosen via the Harr-measure. We numerically analyze our scheme and the influence of imperfect input states caused by non-simultaneous arrival photons.
\end{abstract}


\maketitle


\section{Introduction}
A major goal in quantum information is the implementation of a genuine and universal quantum computer that can solve computational tasks beyond the power of a classical computer. This goal, however, faces huge challenges because current technology can not maintain coherence long enough to perform a full computation, potentially involving a large number of qubits.  Several quantum algorithms, such as Shor's factoring algorithms \cite{Shor:1,Shor:2}, can compute certain problems for which there is no known classical algorithms with polynomial time complexity. However the near term prospects for Shor's factoring algorithm to overturn the central role of the Extended Church Turing thesis \cite{Nielsen:3} in classical computer science are low due to the difficultly of its experimental realization \cite{Martin-Lopez:4}.

Boson sampling provides a different path towards showing the increased power of quantum computation \cite{Aaronson:5}. The complexity of Boson sampling is believed to exceed that of Shor's factoring algorithm.  If a classical algorithm existed to sample the BosonSampling distribution, even in an approximate sense, it would lead the polynomial hierarchy to collapse to the third level, something that is believed not to occur. Experimental boson sampling with up to 6 photons has been reported, demonstrating the theoretical prediction of the output distribution \cite{Broome:6,Spring:7,Tillmann:8,Crespi:9,Carolan2015}.  Temporal encoding has been used to extended the number of scattering modes to 10 with the potential to be extended much further \cite{Motes2014,he:2016}.  In addition, research on sampling from closely rated distributions has extended our knowledge about the nature of boson sampling \cite{Rohde:10,Lund:11,Bentivegna2015}.

Whilst the hardness of boson sampling is its key feature, the hardness of boson sampling requires that the certification of its output is not easy.  This is because boson sampling is in a class of problems which is harder than the class NP (which Shor's algorithm belongs to).  This leaves room for a classically efficient sampling algorithm to fake the boson sampling distribution in a way that is not efficiently detectable.  This has motivated the recent study of different strategies to certify the boson sampling distribution against alternative distributions which attempt to deceive a verifier. The validation test of Aaronson and Arkhipov discriminates the boson sampling distribution against the uniform one within a constant number of samples \cite{Gogolin:12,Aaronson:13,Spagnolo:14}. Another approach based on the tendency for bosons to bunch together in interferometers rules out the classical sampling distribution which uses distinguishable photons as input to the sampler \cite{Carolan:15,Carolan:16}. A so-called mean-field sampler \cite{Tichy:17} invalidates these strategies by replicating the characteristic bosonic bunching behavior, and this type of sampling has been experimentally performed \cite{Crespi2016}. Although this certification scheme was proposed to rule out plausible physical models by observing if some forbidden events occur in the output modes, its effective range is strictly limited to the symmetric situation in which we need to artificially construct a boson sampler with a cyclic-symmetry initial state and an interaction described by a Fourier matrix \cite{Tichy:17}.

In this paper, we describe a certification scheme that can rule out mean-field sampling devices from the true boson sampling without changing the internal working of the device and uses the same distribution of the input transformation matrices as required for true boson sampling. The model we use for a potential boson sampling device is as a black box whose detailed internal functioning is unknown.  The black box takes as an input a Haar-random unitary transformation matrix and an input Fock state described by a sequence of zeros and ones.  The device outputs Fock basis measurement results which are sequences of integers.  In our certification protocol, we inject a test input state consisting of single photons into every mode.   The mean-field sampling devices produce uniform distributions for this test state and true boson sampling does not.  It is then possible to distinguish a boson sampler from a mean-field sampler via analyzing the variations in the output probability distribution.   We also analyze the noise characteristics for partially distinguishable particles in our certification scheme.

\section{Sampling models}
\subsection{Boson Sampling}
An ideal boson sampler scatters $n$ indistinguishable photons by an $N$-port 
interferometer of lossless, linear optical elements, \emph{i.e.} phaseshifters
and beamsplitters, and counts the number of photons exiting from each output
mode. The input state is a Fock state with definite numbers of photons occupying each mode, which is expressed by
\begin{widetext}
\begin{equation}
S_B=\left|s_1,s_2,...,s_N\right\rangle=\prod_{j=1}^N \frac{(\hat{a}
_j^\dagger)^{s_j}}{\sqrt{s_j!}}\left|0\right\rangle = \frac{\left(a_1^\dagger\right)^{s_1} \left(a_2^\dagger\right)^{s_2} \ldots \left(a_N^\dagger\right)^{s_N}}{\sqrt{s_1!s_2!...s_N!}}\left|0\right\rangle,
\end{equation}
\end{widetext}
where $s_j$ represents the number of photons prepared on the $j^{th}$ input mode, $\sum_{j=1}^N s_j=n$, $\hat{a}_j^\dagger$ is the creation operator for the $j_{th}$ input mode and $\left|0\right\rangle$ is the vacuum state. We also define an equivalent alternative notation $\vec{j}=(j_1,j_2,...,j_n)$ for the mode arrangement of input photons, where $j_i$ is the mode occupied by the $i_{th}$ photon and $1 \leq j_i \leq N$. 

The dimensionality of a unitary transformation $\hat{\mathbb{S}}$ between the input and output state vectors will increase exponentially as $n$ and $N$ grow.  However, when restricting the transformations to those of linear optical elements, then the transformation can be represented by an $N \times N$ complex unitary matrix $\Lambda$ which describes the scattering of photons.  One can write this transformation down in terms of the Heisenberg evolution of the creation operators as 
\begin{equation}
	\hat{\mathbb{S}}\hat{a}_j^\dagger\hat{\mathbb{S}}^\dagger= \sum_{k=1}^N \Lambda_{k,j}\hat{a}_k^\dagger.
\end{equation}
which is a linear transformation of the creation operators (and hence the name).  After the linear scattering transformation, a measurement is performed in the Fock basis.  The probability of detecting the event $T_B=\left|t_1,t_2,...,t_N\right\rangle$, whose mode arrangement is $\vec{k}=(k_1,k_2,...,k_n)$, from output ports is
\begin{equation}
P_B(T_B;S_B,\Lambda)=\frac{\left|Permanent(A)\right|^2}{\prod_{i=1}^N s_i!t_i!},\label{ProBS}
\end{equation}
where $A$ is a submatrix of $\Lambda$ with elements $A_{q,p}=\Lambda_{k_q,j_p}$ and the input mode arrangement $\vec{j}$ is defined similarly to the output mode arrangement $\vec{k}$.  The computational task of boson sampling is then defined as producing samples for the output events with the correct probabilities for any given $\Lambda$.

The hardness of boson sampling is shown by using the boson sampling process to make approximations of the permanent of a matrix.  The proof depends on some plausible conjectures about approximating the permanents of i.i.d. Gaussian matrices~\cite{Aaronson:5}. To be close to the Gaussian matrices, it is desirable to choose $\Lambda$ randomly according to the Harr-measure on $N \times N$ unitary matrices and for $N$ to scale at least as $n^2$. Physically, it requires that the scattering process is arbitrarily tunable to perform the corresponding transformation $\Lambda$.  With lossless optical elements and efficient photon detectors, a physically constructed boson sampler would accurately generate samples from this probability distribution.  The existence of a classical algorithm which simulates such a quantum device in polynomial time would imply a collapse in the polynomial hierarchy~\cite{Aaronson:5}.  The collapse of the polynomial hierarchy has quite a technical definition, but essentially a collapse of the hierarchy is akin to a statements of the form $P=NP$ but depending on exactly which level the hierarchy collapses the weaker the statement is about the relationship between $P$ and $NP$.  Nevertheless, it is believed that the hierarchy does not collapse at any level and therefore the boson sampling problem is hard for any classical algorithm.

\subsection{Classical Sampling}
One way to construct a classical analog of boson sampling is to take as an input distinguishable photons instead of indistinguishable photons. Given an input state $S_D=(s_1,s_2,...,s_N)$, the probability to detect the output configuration $T_D=(t_1,t_2,...,t_N)$ is
\begin{equation}
P_D(T_D;S_D,\Lambda)=\frac{Permanent(\Omega)}{\prod_{i=1}^N \sqrt{s_i!t_i!}},\label{ProC}
\end{equation}
where each element in the matrix $\Omega$ is the absolute square of the corresponding element in matrix $A$ as $\Omega_{q,p}=\left|A_{q,p}\right|^2$. Aaronson and Arkhipov showed that sampling from this distribution is possible classically using a polynomial time algorithm \cite{Aaronson:5}.  This result is consistent with the fact that estimating the matrix permanent of matrices with all positive entries can be performed using a classical polynomial time algorithm~\cite{Jerrum:18}.

\begin{figure}[htbp]
\centering
\fbox{\includegraphics[width=8.5cm]{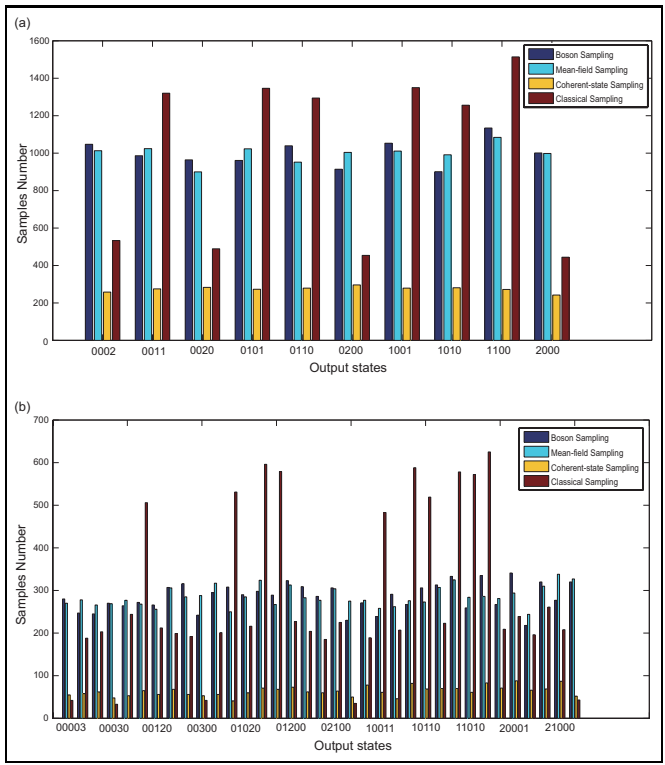}}
\caption{Simulation of boson sampling, mean-field Sampling, coherent-state sampling and classical sampling for (a) two photons with input state $S=(0,1,1,0)$, (b) three photons with input state $S=(0,1,1,1,0)$. Both cases run over 100 Haar-random matrices and sample 100 times for each matrix.}
\label{fig1}
\end{figure}

\begin{widetext}
\subsection{Mean-field Sampling}
The mean-field sampler devised by Tichy~\emph{et.al} \cite{Tichy:17} is an efficiently computable and physically plausible model of a boson sampler imposter. The interference of Bose-Einstein-Condensates motivates the physical implementation of the mean-field sampler, which provides an efficient approximation for the probability distribution of the boson sampling. The input state $S_{Mf}$ is a set of $n$ identical single-photon states
\begin{equation}
\left|\psi\right\rangle=\frac{1}{\sqrt{n}}\sum_{p=1}^n e^{i\theta_p}\hat{a}_{j_p}^\dagger\left|0\right\rangle,\label{eq4}
\end{equation}
where the relative phase $\theta_p$ is randomly chosen from 0 to $2\pi$, the subscript $j_p$ corresponds to the mode arrangement $\vec{j}$ of the boson sampler's input state and denotes the ${j_{p}}_{th}$ input mode. Each single-photon propagates through the linear scattering apparatus and evolves to the state
\begin{eqnarray}
\hat{\mathbb{S}}\left|\psi\right\rangle&=&\frac{1}{\sqrt{n}}\sum_{p=1}^n e^{i\theta_p}\hat{\mathbb{S}}\hat{a}_{j_p}^\dagger\hat{\mathbb{S}}^\dagger\left|0\right\rangle\nonumber\\
&=&\frac{1}{\sqrt{n}}\sum_{k=1}^N \sum_{p=1}^n e^{i\theta_p}\Lambda_{k,j_p}\hat{a}_{j_p}^\dagger\left|0\right\rangle.
\end{eqnarray}
On the side of output ports, each photon would be detected on the $k_{th}$ mode with probability
\begin{equation}
p_k=\frac{1}{n}\left|\sum_{p=1}^n e^{i\theta_p}\Lambda_{k,j_p}\right|^2.
\end{equation}
Injecting $n$ single-photon states as per Eq.~(\ref{eq4}) into the scattering setup shot by shot, we can detect the output event  $T_{Mf}=(t_1,t_2,...,t_N)$, whose mode arrangement is $\vec{k}=(k_1,k_2,...,k_n)$, with probability
\begin{eqnarray}
P_{Mf}(T_{Mf};S_{Mf},\Lambda) &=& \binom{n}{t_1}\binom{n-t_1}{t_2}...\binom{t_N}{t_N}p_1^{t_1}p_2^{t_2}...p_N^{t_N}\nonumber\\
&=& \frac{n!}{n^n\prod_{l=1}^N t_l!}\prod_{k=1}^{N}\left|\sum_{p=1}^n e^{i\theta_p}\Lambda_{k,j_p}\right|^{2t_k}\nonumber\\
&=& \frac{n!}{n^n\prod_{l=1}^N t_l!}\prod_{q=1}^{n}\left|\sum_{p=1}^{n}e^{i\theta_p}\Lambda_{k_q,j_p}\right|^2.\label{ProMf}
\end{eqnarray}
In our case, we supposed the relative phase $\theta_p$ with same subscript $p$ among these single-photon state was identical, but would be randomly regenerated for the next run of sampling. Repeatedly running the sampler time after time, we will get the average probability by integrating $\theta_p$ as
\begin{eqnarray}
\tilde P_{Mf}(T_{Mf};S_{Mf},\Lambda) &=& \int_{0}^{2\pi} \frac{\,d\theta_1}{2\pi}...\int_{0}^{2\pi} \frac{\,d\theta_n}{2\pi}P_{Mf}(T_{Mf};S_{Mf},\Lambda)\nonumber\\
&=& \frac{n!}{n^n\prod_{l=1}^N t_l!}\prod_{q=1}^{n}\sum_{p=1}^{n}\left|\Lambda_{k_q,j_p}\right|^2.\label{ProMf2}
\end{eqnarray}

\subsection{Coherent-state Sampling}
As the name implies, the coherent-state sampler takes the input as a product of coherent states $S_{Cs}=\left|e^{i\chi_1}\left|\alpha_1\right|,e^{i\chi_2}\left|\alpha_2\right|,...,e^{i\chi_N}\left|\alpha_N\right|\right\rangle$, where the $\left|\alpha_j\right|$ is the amplitude of the state and the $\chi_j$ is the relative phase randomly chosen from 0 to $2\pi$. To implement the coherent-state sampler, one only needs a source of coherent light from a laser which is split into $N$ modes which results in the required value of $\alpha_j$.  This is, generally speaking, easier than preparing many indistinguishable single-photon states. After propagating through the linear scattering unitary, the transformation of coherent states is 
\begin{equation}
\hat{\mathbb{S}}S_{Cs} = exp({-\frac{1}{2}\sum_{j=1}^N \left|\alpha_j\right|^2})exp(\sum_{k=1}^{N}\sum_{j=1}^{N} e^{i\chi_j}\alpha_j\Lambda_{k,j}\hat{a}_k^\dagger)\left|0\right\rangle
\end{equation} 
The state of output ports is still a product of coherent states. After renormalization, the final state is 
\begin{eqnarray}
\begin{aligned}
\hat{\mathbb{S}}S_{Cs}
&=\bigotimes_{k=1}^N exp(-\frac{1}{2}\left|\sum_{j=1}^N e^{i\chi_j}\alpha_j\Lambda_{k,j}\right|^2)exp(\sum_{j=1}^N e^{i\chi_j}\alpha_j\Lambda_{k,j}\hat{a}_k^\dagger)\left|0\right\rangle\\
&=\bigotimes_{k=1}^N exp(-\frac{1}{2}\left|\sum_{j=1}^N e^{i\chi_j}\alpha_j\Lambda_{k,j}\right|^2)\sum_{n_k=0}^\infty\frac{(\sum_{j=1}^Ne^{i\chi_j}\alpha_j\Lambda_{k,j})^{n_k}}{\sqrt{n_k!}}\left|n_k\right\rangle.
\end{aligned}
\end{eqnarray}
The probability to detect the output Fock state $T_{Cs}=(t_1,t_2,...,t_N)$ is
\begin{equation}
P_{Cs}(T_{Cs};S_{Cs},\Lambda)
=exp(-\sum_{k=1}^N \left|\sum_{j=1}^N e^{i\chi_j}\alpha_j\Lambda_{k,j}\right|^2)\prod_{k=1}^N \frac{\left|\sum_{j=1}^N e^{i\chi_j}\alpha_j\Lambda_{k,j}\right|^{2t_k}}{t_k!}.\label{ProCs}
\end{equation}
After averaging over the random phase $\chi_j$, the probability becomes
\begin{equation}
\tilde P_{Cs}(T_{Cs};S_{Cs},\Lambda)=exp(-\sum_{k,j=1}^N \left|\alpha_j\Lambda_{k,j}\right|^2)\prod_{k=1}^N \frac{(\sum_{j=1}^N \left|\alpha_j\Lambda_{k,j}\right|^2)^{t_k}}{t_k!}.
\end{equation}

Using the probabilities in Eq.~(\ref{ProBS}),~(\ref{ProC}),~(\ref{ProMf}) and~(\ref{ProCs}), we have simulated boson sampling, classical sampling, mean-field sampling and coherent-state sampling in the case of 2 and 3 photons averaging over 100 Harr-random unitary matrices as depicted in FIG.~\ref{fig1}. The counting statistics of boson sampling is similar to mean-field sampling, while classical sampling obviously behaves higher on coincidence outcomes but lower on those with multiply populated modes. This phenomenon implies that mean-field sampling and coherent-state sampling abide by the bunching tendency of boson sampling. In the simulation, we set the input state of the coherent-state sampler to $\alpha_j=1$ on those modes which would have a single photon input and $\alpha_j=0$ for those modes which would have been vacuum.  As the coherent state has an indefinite number of photons, there are many events output from the coherent-state sampler which are not exactly two-photon or three-photon events in the respective simulations.  The coherent-state sampler therefore yields many events containing more or less photons in total and these events are not shown in FIG.~\ref{fig1}.  When comparing with boson sampling, the counting statistics of the events shown for the coherent-state sampler is consequently lower. 
\end{widetext} 

\section{The Certification Protocol}

\subsection{Certification via the N photon single photon input state}
Since the certification based on bosonic bunching does not distinguish boson sampling and the mean-field sampling, Tichy \emph{et al.} \cite{Tichy:17} proposed a scheme using a symmetric Fourier matrix and a particular initial state with cyclic symmetry. In their scheme, some forbidden outcomes in the boson sampling will occur in the mean-field sampling. However, their scheme has the obvious disadvantage that it must change the sampling method and deviates from the standard boson sampling construction.

Suppose there is a sampling algorithm where the output can be measured and the input is a linear unitary interaction of single photons and vacuum states.  It is assumed that no details of the internal workings of the implementation of the algorithm are known.  Is it possible to identify whether this sampler is a fully quantum boson sampler or a possibly classical mean-field sampler? In this paper, we present numerical evidence that inputting an input test state which is constructed from single photons can solve this problem. The test state of is $S_T=(1,1,...,1)$, whose mode arrangement is $\vec{j}=(1,2,...,N)$.  A correspondingly complex test state for the mean-field sampler is $N$ runs using a single-photon state prepared as $\left|\psi\right\rangle=\frac{1}{\sqrt{N}}\sum_{l=1}^N e^{i\theta_l}\hat{a}_l^\dagger\left|0\right\rangle$. In this case, the sum term in Eq.~(\ref{ProMf2}) becomes $\sum_{l=1}^N \left|\Lambda_{k_q,l}\right|^2$, which is equal to 1 as $\Lambda$ is unitary. Thus, given the test state as the input state of a mean-field sampler, the outcome $T_{Mf}$ will be detected with probability
\begin{equation}
\tilde P_{Mf}(T_{Mf};S_{T},\Lambda)=\frac{n!}{n^n\prod_{k=1}^N t_k!}.\label{PMfAverage}
\end{equation}
From this expression we can see that the probability distribution is not a function of the unitary matrix $\Lambda$ but only a function of the number of photons $n$ and the output configuration $T_{Mf}$.  Since the number of photons is equal to the number of input modes, the probability distribution is determined only by the dimension of the unitary matrix.  The invariance of this distribution with respect to changes in the unitary matrix suggests that there exists an efficient method for discrimination between the mean-field sampler and the fully quantum boson sampler~\cite{Aaronson:13}.  Here we study numerically a scenario with small numbers of photons where the entire probability distribution can be feasibly studied.

\begin{figure}[htbp]
\centering
\fbox{\includegraphics[width=8.5cm]{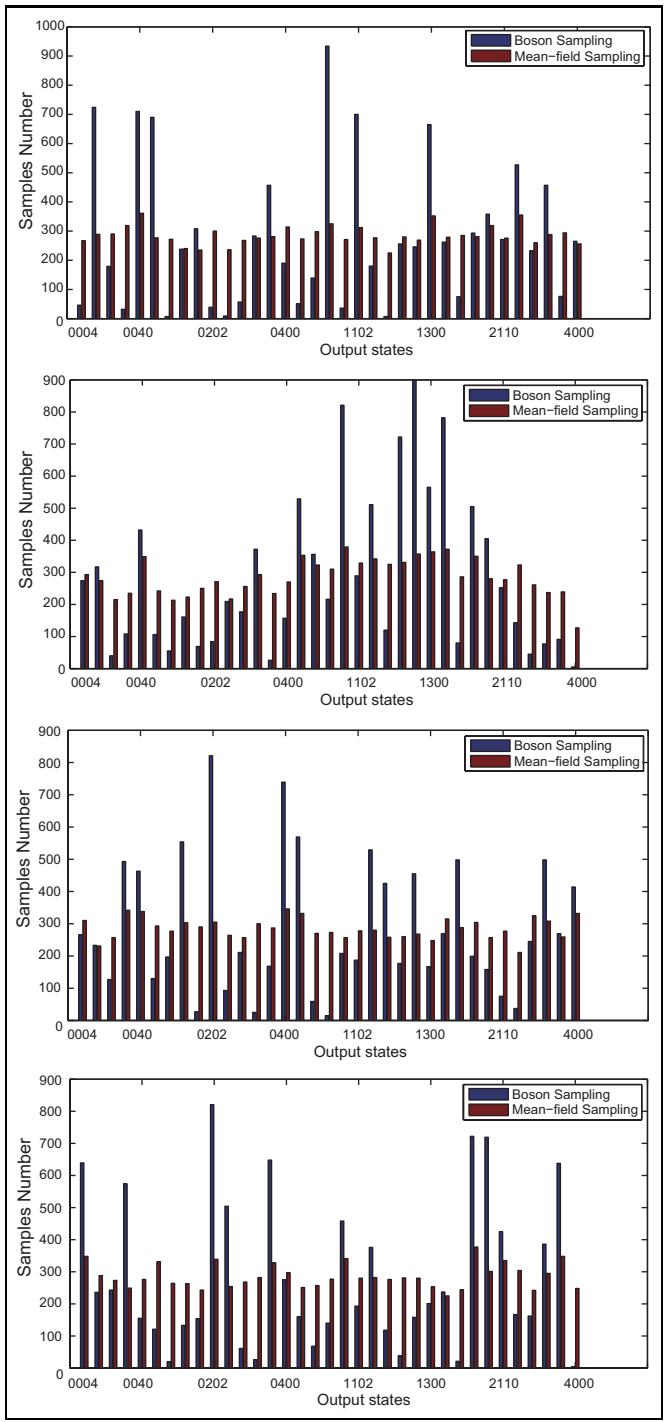}}
\caption{Simulation of boson sampling and mean-field sampling with the test state $S_T=(1,1,1,1)$. Each panel represents a different choice of Haar-random unitary matrix.  For each choice of unitary 10,000 samples are taken.  For this quantity of samples, the maximum standard deviation of the number of samples is 50.  There are cases in each panel for which the number of events sampled deviate between the two cases by 5 to 10 standard deviations.}
\label{fig2}
\end{figure}

In FIG.~\ref{fig2}, we randomly selected four $4 \times 4$ unitary matrices by the Harr-measure and numerically simulated boson sampling and mean-field sampling with the test input state $S_T=(1,1,1,1)$.  The number of samples simulated for all situations is $10000$.  The counting statistics of the mean-field sampling presents as relatively uniform when compared with boson sampling in agreement with the theory above. The sample size here is not large enough for the effect of the randomly varying relative phase $\theta_l$ to completely vanish so the counting statistics of mean-field sampling does not completely satisfy the probability given in Eq.~(\ref{PMfAverage}). Fortunately, the sampling statistic of the mean-field sampler approximates the uniform distribution in the case of an arbitrary unitary matrix, whilst the sampling statistic of the boson sampler will vary for different matrices. For a particular matrix, the difference between the maximum event and the minimum event in boson sampling statistics is larger than those in the relatively uniform mean-field sampling. By using this particular test state as an input, the distributions of boson sampling and mean-field sampling can be distinguished without needing to modify the sampler, even without any prior knowledge about the linear scattering matrix.

\begin{figure}[htbp]
\centering
\fbox{\includegraphics[width=8.5cm]{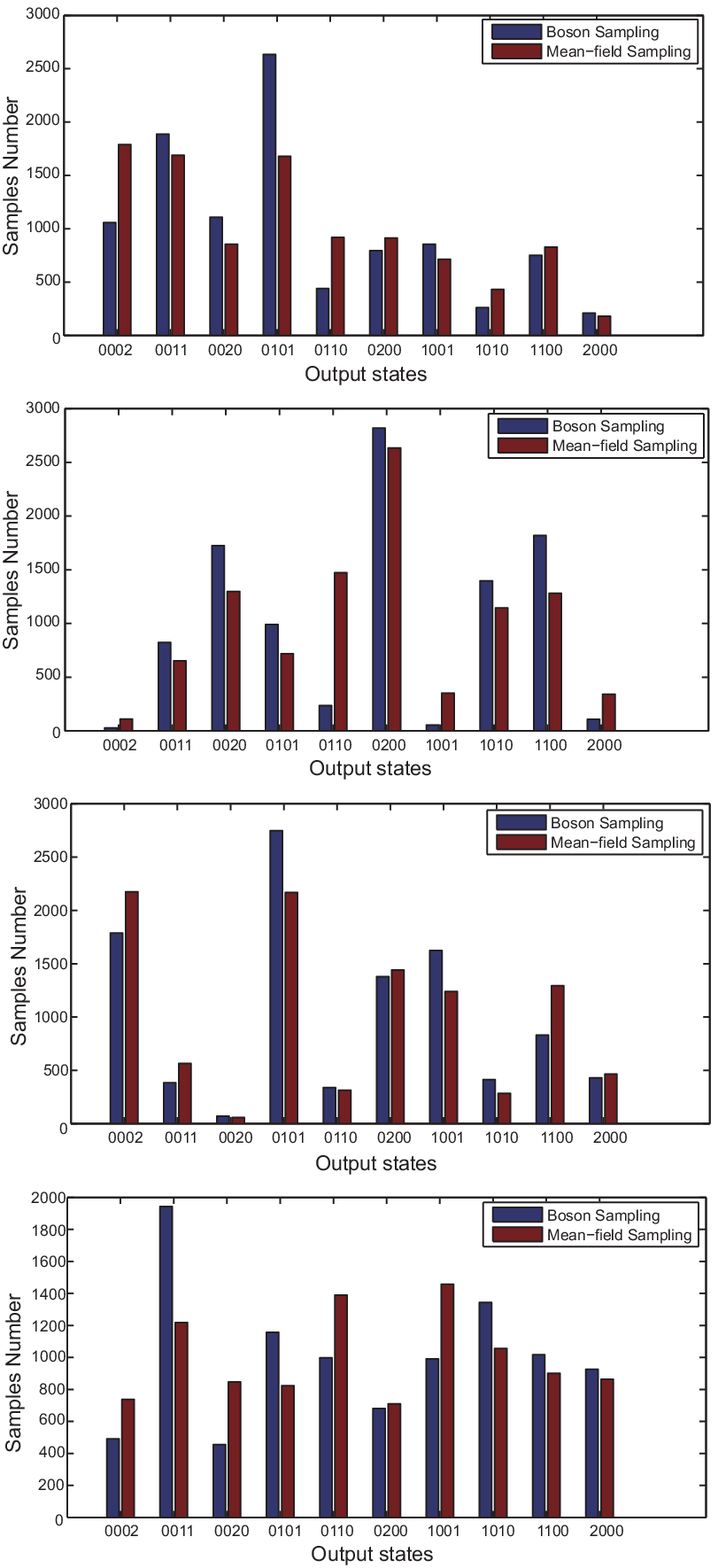}}
\caption{Simulation of boson sampling and mean-field sampling with the input state $S=(0,1,1,0)$. We chose four random matrices by Harr-measure and sampled 10000 times for each matrix.  The same expected variation as FIG.~\ref{fig2} applies to this data.}
\label{fig3}
\end{figure}

In fact, when returning to the case where the input state does not have all modes occupied by a single photon, the most likely event still occurs in the boson sampling distribution and not the mean-field distribution when a large number of samples are taken as shown in FIG.~\ref{fig3}. One can firmly believe that the most sampled event is from the boson-sampling distribution. We conjecture the reason is that random phase of input state for the mean-field sampler limits the maximum value in probability distribution. However, this method may need a number of samples which depends exponentially on the number of input photons and so this may be an inefficient protocol to distinguish the boson sampling and mean-field distributions. 

Our verification protocol can be compared and contrasted with those described in the introduction~\cite{Aaronson:13, Tichy:17}.  One can see that this protocol is different to these previous ones, but has some related structure.  The protocol in~\cite{Aaronson:13} tests if the output distribution is uniformity in the $0$ and $1$ subspace of detections or is a true boson sampling distribution proportional to matrix permanents.  This protocol is efficient and stays completely within the boson sampling model of $n$ photons into $N > n^2$ modes.  It can, however, be efficiently deceived with the classical sampling distribution.  The protocol of~\cite{Tichy:17} deviates from the standard boson sampling model but, given some assumptions about the nature of the linear interaction, is an efficient test of a necessary condition to perform boson sampling.  In particular it finds the ability to perform non-classical interference.   Our protocol also deviates from the standard boson sampling, but in a much simpler way than that in \cite{Tichy:17}.  We add more photons to the input state so that all input modes contain single photons.  In this case, one is now in the regime of \cite{Aaronson:13} where to discriminate the mean-field sampler and boson sampling one must distinguish uniformity from non-uniformity.  When applying our criterion to the distribution which does scatter into $N > n^2$ modes, the distinction between the maximal and minimally sampled events still exists, but as stated above, one may need to gather an exponentially scaling number of samples to distinguish these distributions.  Depending on the cost of single photons relative to modes, it may be desirable to work in the small photon number regime and take an exponential number of samples to perform our test instead of using the $N=n$ case.  It also must be remembered that none of these schemes {\em efficiently} verify the device's proper functioning as a boson sampler as this is believed to not be possible~\cite{Aaronson:5}.

\subsection{Robustness for the partially distinguishable boson sampling}     
Boson sampling in practice is expected to be influenced by experimental inaccuracies in the inputting, scattering and detecting of single photons. We now consider the case of imprecision in preparing input states.  In particular, we consider inputs states to the boson sampler constructed from partially distinguishable photon states.  In this analysis, the degree of freedom used to induce distinguishability is the mutual delay of injected photons.  The theory we use to describe this effect considers orthogonal spatio-temporal modes.  By performing a Gram-Schmidt orthogonalization on the spatio-temporal degrees of freedom of the input state, the single particle in each mode can be represented as
\begin{equation}
\hat{a}_{j_r}^\dagger=\sum_{k=1}^{j_r}C_{r,k}\hat{a}_{j_r,t_k}^\dagger,
\end{equation}
where the coefficients satisfy normalization such that $\sum_{k=1}^{j_r}\left|C_{j_r,k}\right|^2=1$ and the single-photon states are orthonormalized as $\left\langle0\right|\hat{a}_{p,t_m}\hat{a}_{q,t_n}^\dagger\left|0\right\rangle=\delta_{p,q}\delta_{m,n}$.  Here the double subscript represent spatial and temporal modes labels respectively. In this notation we consider the temporal degree of freedom to form an orthonormal basis of arriving time $\{\left|t_1\right\rangle, \left|t_2\right\rangle, ..., \left|t_{j_r}\right\rangle\}$, the input state can be expressed as \cite{Tichy:17,Tichy:19}
\begin{equation}
	\label{badstate}
	|S_B\rangle=\hat{a}_{j_1,t_1}^\dagger\sum_{k_2=1}^2\sum_{k_3=1}^3...\sum_{k_n=1}^n\prod_{r=2}^n C_{r,k_r}\hat{a}_{j_r,t_{k_r}}^\dagger\left|0\right\rangle.
\end{equation}
This equation can be qualitatively understood as mode 1 defining a basis into which all other modes are compared, so the first mode is the ``pure'' mode.  The second mode then has some common overlap with the first mode and hence can be divided up into two components which overlaps with the first mode and some other orthogonal component which results in the sum over the modes only being required up to two.  The third mode in which a photon is created will then have some component which overlaps with the first and second modes and then an orthogonal component which means there will be three terms in this sum.  And so on in this manner until all photons have been expanded.  The terms which have identical subscript $t_{k_r}$ describes an input state of temporally indistinguishable photons, which relates to a standard boson sampling. The term with different subscript $t_{k_r}$ corresponds to a classical sampling with distinguishable particles. The complex case is terms with some identical and some different $t_{k_r}$, which relate to the partially distinguishable boson sampling. Adjusting the coefficient $C_{r,k_r}$, the input state gives rise to a sampling in the range from fully indistinguishable to fully distinguishable particles. 

\begin{figure}[htbp]
\centering
\fbox{\includegraphics[width=8.5cm]{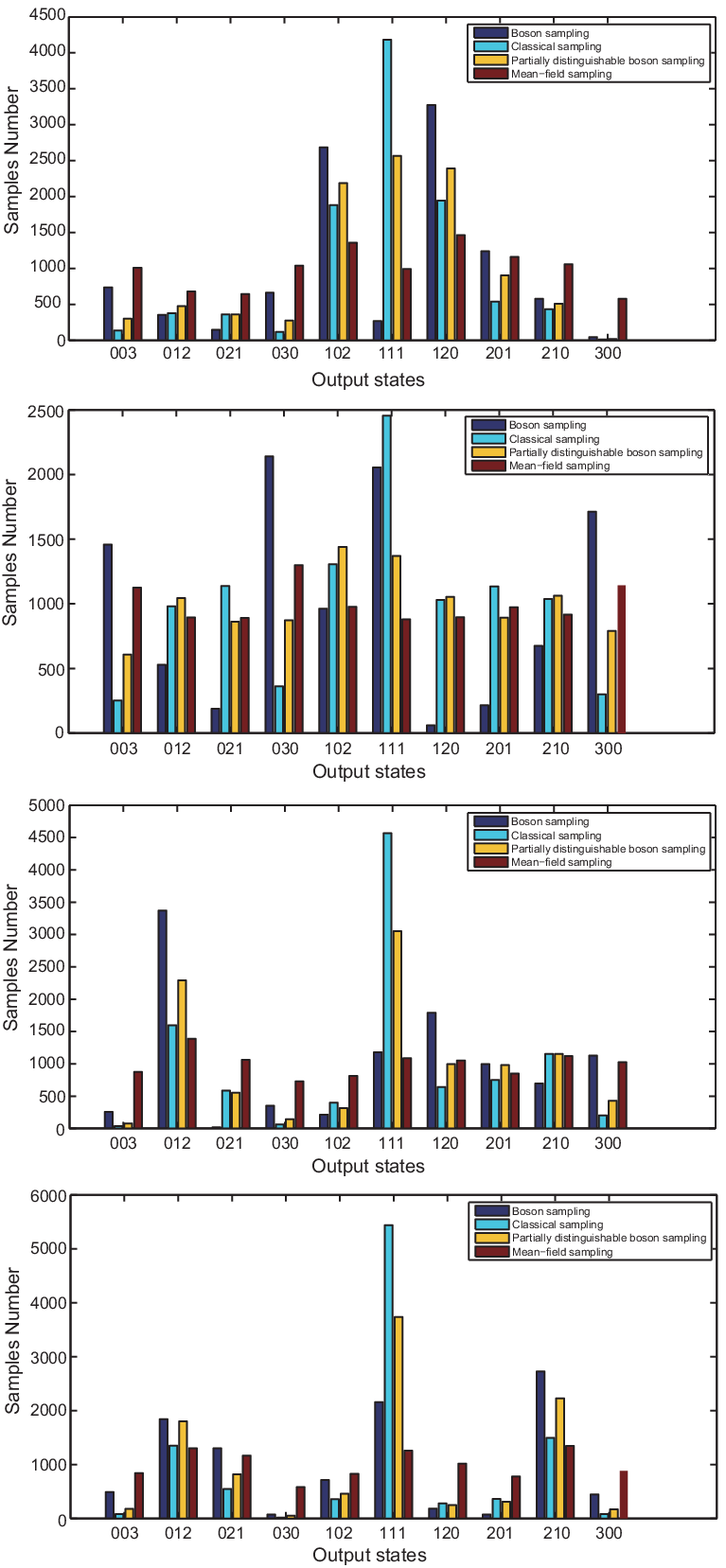}}
\caption{Simulation of boson sampling, classical sampling, partially distinguishable boson sampling and mean-field sampling with the input state $S=(1,1,1)$.  Four randomly chosen matrices are displayed with 10,000 samples for each case.  We chose the coefficients for the Fock basis state in Eq.~(\ref{badstate}) to be identical, \emph{i.e.} $\frac{1}{\sqrt{6}}$ as there are 6 Fock basis states in this example.  In terms of two-photon interference, this is equivalent to having a 50\% probability of finding that the input particles were indistinguishable.}
\label{fig4}
\end{figure}

By injecting the test state $S_T=(1,1...1)$, we still can differentiate the mean-field sampling from the partially distinguishable boson sampling. As shown in FIG.~\ref{fig4}, we simulate the partially distinguishable boson sampling and the mean-field sampling with the test state $S_T=(1,1,1)$. For the partially distinguishable boson sampling, the fluctuations vary considerably depending on random matrices, while mean-field sampling presents a stable distribution.

\section{Conclusions}
Boson sampling with identical bosons provides evidence that quantum computers can outperform classical computers.  However its verification is known to be difficult due to its computational complexity being outside the class $NP$. In our paper, we have proposed a certification scheme for the specific case of discriminating the boson sampler from the mean-field sampler. Without any limitation on the scattering matrices, the protocol only needs to prepare a test state and primitively analyze the output probability distribution. We have numerically simulated samples of the boson sampling and the mean-field sampling distributions when inputting the test state. For the output probability distributions, the statistics of boson sampling fluctuate more significantly than mean-field sampling. Considering the influence of imperfect input states, we have shown the certification capability of the protocol for partially distinguishable boson sampling against mean-field sampling.

In summary, we describe an experimental approach to verify the actual boson sampling against a physically plausible imposter. Compared to previous methods, our scheme retains a scattering matrix randomly chosen by the Harr-measure and continues to work under the noise of partial distinguishability caused by non-simultaneously arriving photons. This work does not address the efficiency problem of our scheme. Optimizing the scheme to reduce the number of samples needed will be the direction of future work. 

\section*{Acknowledgements}
National Natural Science Foundation of China (11475160, 61575180); The Natural Science Foundation of Shandong Province (ZR2014AQ026, ZR2014AM023); The Fundamental Research Funds for the central universities of China (201313012); The State Scholarship Fund of China Scholarship Council (201306330035); The special fund for National Key Laboratory of Electromagnetic Environment (201500005); Australian Research Council Centre of Excellence for Quantum Computation and Communication Technology (CE110001027).

\bibliography{bosonsampler}

 

%

\end{document}